\documentclass[preprint,prd,aps,showpacs,showkeys,nofootinbib]{revtex4}
\usepackage{graphicx}
\usepackage{dcolumn}
\usepackage{bm}
\begin{document}

\title{Single production of the doubly charged scalar in the littlest Higgs model}
\author{Chong-Xing Yue, Shuang Zhao, Wei Ma}
\affiliation{Department of Physics, Liaoning Normal University, Dalian
116029, China}
\date{\today}

\thanks{E-mail:cxyue@lnnu.edu.cn}

\begin{abstract}
Single production of the doubly charged scalars
$\Phi^{\pm\pm}$ via $e\gamma$, $ep$ and $pp$ collisions is studied
in the context of the little Higgs$(LH)$ model. Our numerical
results show that the new particles $\Phi^{\pm\pm}$ can be
abundantly produced and their possible signatures should be detected
in future high energy linear $e^{+}e^{-}$ collider$(ILC)$. The cross
section for single production of $\Phi^{\pm\pm}$ at the $LHC$ is
much smaller than that at the $ILC$ or the $THERA$.
\end{abstract}

\maketitle

\vspace{0.8cm}
\newpage
\section*{I. Introduction}
\hspace{5mm}It is widely believed that the mechanism of electroweak
symmetry breaking$(EWSB)$ and the origin of the particle mass remain
prominent mystery in current particle physics in spite of the
success of the standard model$(SM)$ tested by high energy
experimental data. There has been no experimental evidence of the
$SM$ Higgs boson existing. Furthermore, the neutrino oscillation
experiments have made one believe that neutrinos are massive,
oscillate in flavor, which presently provides the only experimental
hints of new physics$(NP)$ [1]. Thus, the $SM$ can only be an
effective theory below some high energy scales. Other $EWSB$
mechanisms and extended Higgs sectors have not been excluded in the
theoretical point of view.

Doubly charged scalars arise in many new physics models with
extended Higgs sectors, such as left-right symmetric models [2],
Higgs triplet models [3], 3-3-1 models [4], and little Higgs models
[5]. In these new physics models, doubly charged scalars appear
typically as components of the $SU(2)$ triplet representations,
which do not couple to quarks and their couplings to leptons break
the lepton number by two units. As a result, these new scalar
particles have a distinct experimental signature, namely a same sign
pair of leptons. Thus, discovery of a doubly charged scalar particle
in future high energy colliders would be a definite signal of $NP$
beyond the $SM$, which would help us to understand the Higgs sector
and more importantly what lies beyond the $SM$.

 Since charge conservation prevents doubly charged scalars from
 decaying to a pair of quarks and their Yukawa couplings violate
 lepton number conservation, this kind of new particles would have a
 distinct experimental signature, which has lower backgrounds. In
 addition, the presence of doubly charged scalars provides a simple
 explanation to the lightness of neutrinos via the see-saw mechanism [6],
  contents with recent data on neutrino oscillations [1]. This fact
 has lead to many studies involving production and decay of doubly
 charged scalars at present or future high energy collider
 experiments within specific popular models beyond the $SM$. For
 example, studies of production for doubly charged scalars have been
 given in the context of some specific popular models at hadron colliders [7,8]
 and different modes of
 operation of lepton colliders such as $e^{+}e^{-}$ mode [9],
 $e^{-}e^{-}$ mode [10], and $e\gamma$ or $\gamma\gamma$ mode [11].

 Although there are lot of works on production and decays of doubly charged
 scalars in the literature, it is need to be further studied in the context
 of the little Higgs models. There are several motivations to perform this study.
 First, little Higgs theory can be seen as one of the important candidates of the
 $NP$ beyond the $SM$, most of the little Higgs models predict the
 existence of the doubly charged scalars. However, in previous works
 on studying the phenomenology of the little Higgs models, studies
 about doubly charged scalars are very little. Second, three types of colliders
 related to particle physics research seem
 to be promising in the next decade. Namely, they are the $CERN$ Large
 Hadron Collider$(LHC)$, the high energy linear $e^{+}e^{-}$ collider$(ILC)$,
 and the linear-ring type $ep$ collider $(LC\bigotimes LHC)$
 called $THERA$. The phenomenology of the new physics models should be
 analyzed taking
 into account all three types of colliders. So far, a complete study on single
 production of the doubly charged scalars has not been presented in the context of
 the little Higgs models. Third, studying the possible signals of the doubly
 charged scalars in future high energy colliders can help the collider experiments to
 test little Higgs models, distinguish different $NP$ models, and further to
 probe the production mechanism of the neutrino mass. Thus, in this paper, we will
 consider single production of the doubly charged scalars
 $\Phi^{\pm\pm}$ predicted by the littlest Higgs$(LH)$ model [12]
 and see whether their signals can be detected in future three types
 of high energy colliders $(ILC$,   $THERA$,   and  $LHC)$.

  There are several variations of the little Higgs models, which
  differ in the assumed higher symmetry and in the representations
  of the scalar multiples. In matter content, the $LH$ model is the
  most economical little Higgs model discussed in the literature,
  which has almost all of the essential feature of the little Higgs
  models. The severe constraints on its free parameters coming from
  the electroweak precision data can be solved by introducing the
  T-parity [13]. Using of the fact that the $LH$ model contains a
  complex triplet scalar $\Phi$, Refs.[14,15] have discussed the
  possibility to introduce lepton number violating interactions and generation
  of neutrino mass in the framework of the $LH$ model. It has been
  shown [16] that the neutrino masses can be given by the term
  $\nu'Y_{ij}$, in which $\nu'$ is the vacuum expectation value$(VEV)$ of
  the complex triplet scalar $\Phi$ and $Y_{ij}$ is its
  Yukawa coupling constant. As long as the $ VEV$ $\nu'$ is restricted
  to be extremely small, the value of $Y_{ij}$ is of naturel order
  one. So, the doubly charged scalars $\Phi^{\pm\pm}$ predicted by
  the $LH$ model might produce large contributions to some of lepton
  flavor violating$(LFV)$ processes [17,18].

  Using the current experimental upper limits on the branching
  ratios $Br(l_{i}\rightarrow l_{j}\gamma)$ and $Br(l_{i}\rightarrow
  l_{j}l_{k}l_{k})$, Ref.[18] obtains constraints on
  the relevant free parameters of the $LH$ model. Taking into account these
  constraints, we further consider
  the contributions of $\Phi^{\pm\pm}$ to the $LFV$ processes $e^{\pm}e^{\pm}
  \rightarrow l_{i}^{\pm}
  l_{j}^{\pm}$ and $e^{+}e^{-}\rightarrow l_{i}^{\pm} l_{j}^{\pm }$$(l_{i}$ or $l_{j}\neq
  e)$ in Ref.[18]. Based on Ref.[18], in this paper, we
  will discuss all possible processes for single production of the doubly charged scalars
  $\Phi^{\pm\pm}$ in the future high energy collider experiments.
  The rest of this paper is organized as follows. The relevant
  couplings of the doubly charged scalars $\Phi^{\pm\pm}$ are
  summarized in Section II. Sections III , IV, and V are
  devoted to the computation of the cross sections for single production of the
  doubly charged scalars $\Phi^{\pm\pm}$ at the $ILC$, $THERA$, and
  $LHC$, respectively. Some phenomenological analysis are also
  included in the three sections. Our conclusions and simple discussions
  are given in section VI.

\section*{II. The relevant couplings of the doubly charged scalars
$\Phi^{\pm\pm}$}

\hspace{5mm}The $LH$ model [3] consists of a non-linear $\sigma$
model with a global $SU(5)$ symmetry and a locally gauged symmetry
$[SU(2) \times U(1)]^{2}$. The global
 $SU(5)$ symmetry is broken down to its subgroup $SO(5)$ at a scale
 $f\sim TeV$, which results in 14 Goldstone bosons$(GB's)$. Four of
 these $GB's$ are eaten by the new gauge bosons $(W_{H}^{\pm}, Z_{H},
 B_{H})$, resulting from the breaking of $[SU(2) \times
 U(1)]^{2}$, giving them masses. The Higgs boson remains as a light
 pseudo-Goldstone boson, and other $GB's$ give masses to the $SM$
 gauge bosons and form the complex triplet scalar $\Phi$. Thus, the
 remaining ten $GB's$ can be parameterized as:

 \begin{equation}
\Pi=\left( \begin{array}{ccc}0&\frac{H^{+}}{\sqrt{2}}&\Phi^{+}\\
\frac{H}{\sqrt{2}}&0&\frac{H^{*}}{\sqrt{2}}\\\Phi&\frac{H^{T}}{\sqrt{2}}
&0\end{array}\right) \hspace{5mm}with \hspace{5mm} \Phi=\left(
\begin{array}{cc}\Phi^{++}&
\frac{\Phi^{+}}{\sqrt{2}}\\
\frac{\Phi^{+}}{\sqrt{2}}&\frac{\Phi^{0}+i\Phi^{p}}{\sqrt{2}}\end{array}\right).
\end{equation}

Where H is the $SM$ Higgs doublet, $\Phi^{++}, \Phi^{+}, \Phi^{0}$,
 and $\Phi^{p}$ are the components of the complex triplet scalar
 $\Phi$ , which are degenerate at lowest order with a common mass
 $M_{\Phi}$.
 The couplings of the doubly charged scalar $\Phi^{--}$ to other
 particles, which are related to our calculation, can be written as [15,19]:
\begin{eqnarray}
\Phi^{--}W_{\mu}^{+}W_{\nu}^{+}&:&2i\frac{e^{2}}{S_{W}^{2}}\nu'g_{\mu\nu},
\\
\Phi^{--}W_{H\mu}^{+}W_{H\nu}^{+}&:&2i\frac{e^{2}}{S_{W}^{2}}\frac{(c^{4}+
s^{4})}{2s^{2}c^{2}}\nu'g_{\mu\nu},
\\
\Phi^{--}W_{\mu}^{+}W_{H\nu}^{+}&:&-2i\frac{e^{2}}{S_{W}^{2}}\frac{(c^{2}-
s^{2})}{2sc}\nu'g_{\mu\nu},
\\
\gamma\Phi^{++}\Phi^{--}&:& -2ie(p_{1}-p_{2})_{\mu},
\\
\Phi^{--}l_{i}^{+}l_{j}^{+}&:& 2iY_{ij}^{*}CP_{R},
\\
\Phi^{--}W_{\mu}^{+}\Phi^{+}&:&-i\frac{e}{S_{W}}(p_{1}-p_{2})_{\mu},
\\
\Phi^{--}\Phi^{+}W_{H\mu}^{+}&:&i\frac{e}{S_{W}}\frac{(c^{2}-s^{2})}{2sc}
(p_{1}-p_{2})_{\mu}.
\end{eqnarray}
Where $S_{W}=\sin\theta_{W}$, $\theta_{W}$ is the Weinberg angle.
The free parameter c $(s=\sqrt{1-c^{2}})$ is the mixing parameter
between $SU(2)_{1}$ and $SU(2)_{2}$ gauge bosons.
$P_{R}=(1+\gamma_{5})/2$ is the right-handed projection operator and
C is the charge conjugation operator. The indices $i$ and $j$
represent three generation leptons $e$, $\mu$, or $\tau$. Eq.(6)
gives the flavor-diagonal$(FD)$ couplings and the
flavor-mixing$(FX)$ couplings of the doubly charged scalar
$\Phi^{--}$ to leptons for $i=j$ and $i\neq j$, respectively.

In the $LH$ model, the neutrino mass matrix can be written as
$M_{ij}=Y_{ij}\nu'$ [15,16]. Considering the current bounds on the
neutrino mass [19], we should have $Y_{ij}\nu'\sim10^{-10}$GeV. If
we assume that the value of the triplet scalar $VEV$ $\nu'$ is
smaller than $1\times10^{-5}$GeV, then there is the Yukawa coupling
constant $Y_{ij}>1\times10^{-5}$, which does not conflict with the
most severe constraints on the $LH$ model from the $LFV$ process
$\mu\rightarrow eee$ [18]. In this case, the decay channel
$\Phi^{--}\rightarrow W^{-}W^{-}$ can be neglected. As a result, the
total decay width $\Gamma_{\Phi}$ of the doubly charged scalar
$\Phi^{--}$ can be approximately written as [15,18]:
\begin{equation}
\Gamma_{\Phi}\approx\frac{3M_{\Phi}Y_{ii}^{2}}{8\pi}.
\end{equation}

In the following sections, we will assume
$Y_{ij}>1\times10^{-5}$GeV, which might produce large contributions
to some $LFV$ processes, and take $M_{\Phi}$ and $Y_{ij}$ as free
parameters to calculate the single production cross sections of the
doubly charged scalar $\Phi^{--}$ in future three types of high
energy collider experiments.
\section*{III. Single production of $\Phi^{--}$ at the $\emph{\textbf{ILC}}$}

\hspace{5mm}It is widely believed that the hadron colliders, such as
$Tevatron$
 and $LHC$, can directly probe possible $NP$ beyond the $SM$ up to a
 few TeV, while the TeV energy linear $e^{+}e^{-}$ collider$(ILC)$
 is also need to complement the probe of the new particles with
 detailed measurement [21]. An unique feature of the $ILC$ is that it
 can be transformed to $\gamma\gamma$ or $e\gamma$ collision with
 the photon beam generated by laser-scatting method. The
 effective luminosity and energy of the $\gamma\gamma$ and $e\gamma$
 collisions are expected to be comparable to
 those of the $ILC$. In some scenarios, they are the best
 instrument for discovery of the $NP$ signatures. The $e\gamma$ collision can
 produce particles, which are kinematically not accessible in the
 $e^{+}e^{-}$ collision at the same collider. For example, for the process
 $e\gamma\rightarrow AB$ with light particle $A$ and new particle $B$,
 the discovery limits can be much higher than in other reactions.

\begin{figure}[t]
\setlength{\unitlength}{1mm}
\begin{center}
\begin{picture}(0,40)(0,0)
\put(-82,-140){\includegraphics{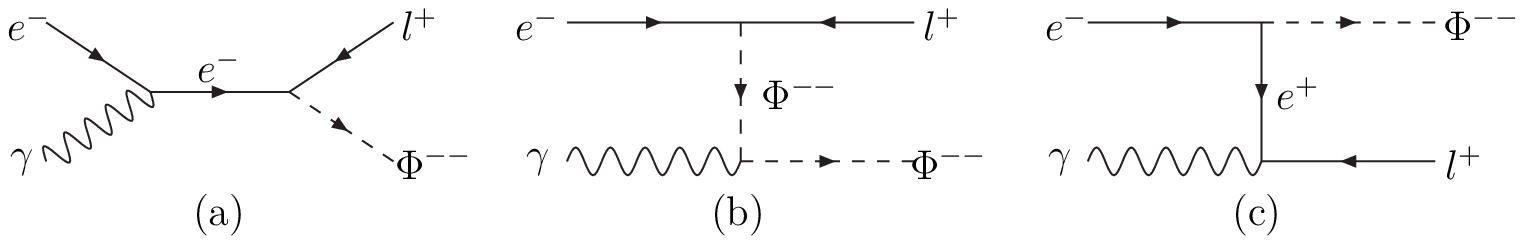}}
\end{picture}
\caption[]{Feynman diagrams for the process $e^{-}\gamma\rightarrow l^{+}\Phi^{--}$.}
\label{fig1}
\end{center}
\end{figure}

 From discussions given in section II, we can see that the doubly charged scalar
 $\Phi^{--}$ can be produced via $e^{-}\gamma$ collision associated with
 a lepton. The relevant Feynman diagrams are shown in Fig.1, in
 which $l^{+}$ is the lepton $e^{+}$, $\mu^{+}$, or $\tau^{+}$.

For the process $e^{-}(P_{1})+\gamma(k_{1})\rightarrow
 l^{+}(P_{2})+\Phi^{--}(k_{2})$, the renormalization amplitude can be written as:
\begin{eqnarray}
M&=&-\frac{eY_{el}}{(P_{1}+k_{1})^{2}}v^{T}(P_{2})C^{-1}P_{R}(\not{\hspace*{-0.15cm}P_{1}+
\not{\hspace*{-0.1cm}k_{1}}})\gamma^{\mu}u(P_{1})\varepsilon_{\mu}(k_{1})\nonumber\\
&&-\frac{2eY_{el}}{(k_{1}-k_{2})^{2}-M^{2}_{\Phi}}v^{T}(P_{2})C^{-1}P_{R}
u(P_{1})(2k_{2}-k_{1})^{\mu}\varepsilon_{\mu}(k_{1})\nonumber\\
&&-\frac{eY_{el}}{(P_{1}-k_{2})^{2}}v^{T}(P_{2})\gamma^{\mu}(\not{\hspace*{-0.15cm}
P_{1}-\not{\hspace*{-0.1cm}k_{2}}})C^{-1}P_{R}u(P_{1})\varepsilon_{\mu}(k_{1})
\end{eqnarray}

After calculating the cross section $\widehat{\sigma}(\widehat{s})$
 for the subprocess $e^{-}\gamma\rightarrow l^{+}\Phi^{--}$, the
 effective cross section $\sigma_{1}(s)$ at the $ILC$ with the
 center-of-mass $\sqrt{s}=2$TeV can be obtained by folding the cross section
 $\widehat{\sigma}_{1}(\widehat{s})$
 with the photon distribution function $f_{\gamma/e}$[22]:
 \begin{equation}
\sigma_{1}(s)=\int_{M_{\Phi}^2/s}^{0.83}dx\widehat{\sigma}_{1}(\widehat{s})
f_{\gamma/e}(x),
\end{equation}
where $x=\widehat{s}/s$, in which $\sqrt{\widehat{s}}$ is the
center-of-mass energy of the subprocess $e^{-}\gamma\rightarrow
l^{+}\Phi^{--}$.

\begin{figure}[t]
\setlength{\unitlength}{1mm}
\begin{center}
\begin{picture}(0,60)(0,0)
\put(-42,-0){\includegraphics{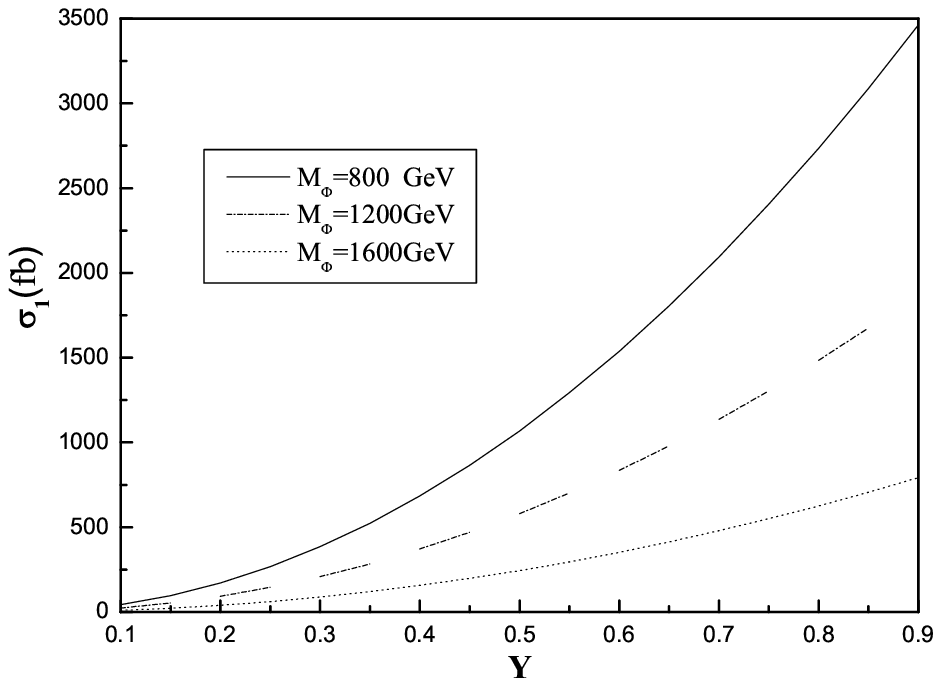}}
\end{picture}
\caption[]{The cross section $\sigma_{1}(s)$ as a function of the $FD$
coupling constant $Y$ for three values of the mass
$M_{\Phi}$.}
\label{fig2}
\end{center}
\end{figure}

The effective cross section $\sigma_{1}(s)$ for the process
$e^{-}\gamma\rightarrow e^{+}\Phi^{--}$ is plotted in Fig.2 as a
function of the FD coupling constant $Y=Y_{ee}$ for $\sqrt{s}=2$TeV
and three values of the $\Phi^{--}$ mass $M_{\Phi}$. From this
figure, one can see that the doubly charged scalar $\Phi^{--}$ can
be abundantly produced via $e\gamma$ collision at the $ILC$
experiment with $\sqrt{s}=2TeV$. For example, for $0.1\leq Y\leq0.9$
and $800GeV \leq M_{\Phi} \leq 1600GeV$, the value of
$\sigma_{1}(s)$ is in the range of $42.8fb\sim3474fb$. If we assume
that the future $ILC$ experiment with $\sqrt{s}=2TeV$ has a yearly
integrated luminosity of $\pounds=500fb^{-1}$, then there will be
ten thousands up to several millions of the doubly charged scalar
$\Phi^{--}$ associated with a positive electron $e^{+}$ to be
generated per year.

In general, the doubly charged scalar $\Phi^{--}$ can decay to the
modes $\Phi^{-}\Phi^{-}$, $W^{-}\Phi^{-}$, $W^{-}W^{-}$ and
$l_{i}^{-}l_{j}^{-}$. However, degeneracy among the components of
the triplet scalar $\Phi$ forbids appearance of the first two modes.
Furthermore, the decay width $\Gamma(\Phi^{--}\rightarrow
W^{-}W^{-})$ is controlled by the triplet scalar $VEV$ $\nu'$. It
has been shown that, for $\nu'<1\times10^{-5}$GeV and
$M_{\Phi}<2000$GeV, $\Phi^{--}$ mainly decays to
$l_{i}^{-}l_{j}^{-}$[10]. In this case, the signatures of
$\Phi^{--}$ is the same-sign lepton pair, including lepton-number
violating final states, which is the $SM$ background free and has
been investigated in Ref.[11]. Thus, as long as the doubly charged
scalar $\Phi^{--}$ is not too heavy, its possible signals should be
detected in future $ILC$ experiments.

The doubly charged scalar $\Phi^{--}$ can also be singly produced
via the $LFV$ processes $e^{-}\gamma\rightarrow l^{+}\Phi^{--}(l\neq
e)$ induced by the lepton number violating interactions. However,
the experimental upper limits on the $LFV$ processes
$\tau\rightarrow eee$ and $\mu\rightarrow eee$ can give severe
constraints on the combination
$|Y_{ii}Y_{ij}|^{2}/M_{\Phi}^{4}(i\neq j)$ [18], which make that the
production cross sections of the $LFV$ processes
$e^{-}\gamma\rightarrow l^{+}\Phi^{--}(l\neq e)$ are very small. For
example, for $Y=0.1$ and $M_{\Phi}=1000$GeV, the effective cross
section for the process $e^{-}\gamma\rightarrow \mu^{+}\Phi^{--}$ is
smaller than $6.9\times10^{-4}fb$. Thus, it is very difficult to
detect the signals of $\Phi^{--}$ via the $LFV$ processes
$e^{-}\gamma\rightarrow l^{+}\Phi^{--}(l\neq e)$ in future $ILC$
experiments.

\section*{IV. Single production of the doubly charged scalar $\Phi^{--}$
at the $\emph{\textbf{THERA}}$}

\hspace{5mm}Although the linear-ring-type $ep$ collider
$(LC\bigotimes LHC)$ with the center-of-mass energy
$\sqrt{s}=3.7$TeV and with the integral luminosity
$\pounds=100pb^{-1}$ has a lower luminosity, it can provide better
condition for studying a lot of phenomena, compared to the $ILC$ due
to the high center-of-mass energy and compared to the $LHC$ due to
clearer environment [23]. Thus, it can be used to detect the
possible signals of some new particles. In this section, we consider
single production of the doubly charged scalar $\Phi^{--}$ predicted
by the $LH$ model via $ep$ collision at the $THERA$.

The doubly charged scalar $\Phi^{--}$ can be singly produced at the
 $THERA$ via the processes:
\begin{equation}
e^{-}p\rightarrow e^{+}X\Phi^{--},\hspace*{0.3cm} e^{-}p\rightarrow
\mu^{+}X\Phi^{--},\hspace*{0.3cm} e^{-}p\rightarrow
\tau^{+}X\Phi^{--},
\end{equation}
which were first studied by Ref.[24]. In terms of observability, the
second and third processes are induced by the lepton number
violating interactions. The corresponding subprocesses can be
unitive written as $e^{-}\gamma\rightarrow l^{+}\Phi^{--}$ $(l=e$,
$\mu$ or $\tau)$. The relevant Feynman diagrams are plotted in
Fig.3. Since the production cross section for the process
$e^{-}\gamma\rightarrow \mu^{+}\Phi^{--}$ or $e^{-}\gamma\rightarrow
\tau^{+}\Phi^{--}$ is much smaller than that of the process
$e^{-}\gamma\rightarrow e^{+}\Phi^{--}$, we will use the equivalent
photon approximation$(EPA)$ approach [25,26] to only consider the
process $e^{-}p\rightarrow e^{+}X\Phi^{--}$ in this section.
\begin{figure}[t]
\setlength{\unitlength}{1mm}
\begin{center}
\begin{picture}(0,40)(0,0)
\put(-82,-150){\includegraphics{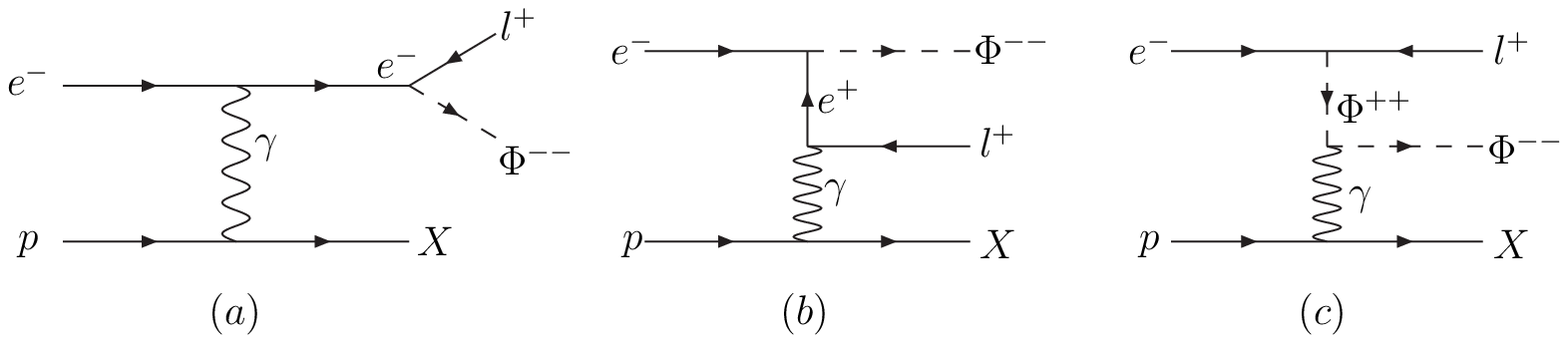}}
\end{picture}
\caption[]{Feynman diagrams for single production of
$\Phi^{--}$ at the $THERA$.}
\label{fig3}
\end{center}
\end{figure}

Using the $EPA$ method, the effective cross section $\sigma_{2}(s)$
at the $THERA$ with $\sqrt{s}=3.7$TeV and $\pounds=100pb^{-1}$ can
be folding the cross section $\widehat{\sigma}_{2}(\widehat{s})$ for
the subprocess $e^{-}\gamma\rightarrow e^{+}\Phi^{--}$ with the
photon distribution function $f_{\gamma/p}(x,\widehat{s})$:
\begin{equation}
\sigma_{2}(s)=\int_{M_{\Phi}^{2}/s}^{(1-m/\sqrt{s})^{2}}dx
\widehat{\sigma}(xs)f_{\gamma/p}(x,xs),
\end{equation}
with $x=\widehat{s}/s$, $m$ is the proton mass and
\begin{equation}
f_{\gamma/p}(x,xs)=f_{\gamma/p}^{el}(x)+f_{\gamma/p}^{inel}(x,xs).
\end{equation}
Where $f_{\gamma/p}^{el}(x)$ and $f_{\gamma/p}^{inel}(x,xs)$ are the
elastic and inelastic components of the equivalent photon
distribution of the proton, which has been extensively studied in
Refs.[25, 26, 27].

\begin{figure}[t]
\setlength{\unitlength}{1mm}
\begin{center}
\begin{picture}(0,60)(0,0)
\put(-42,-0){\includegraphics{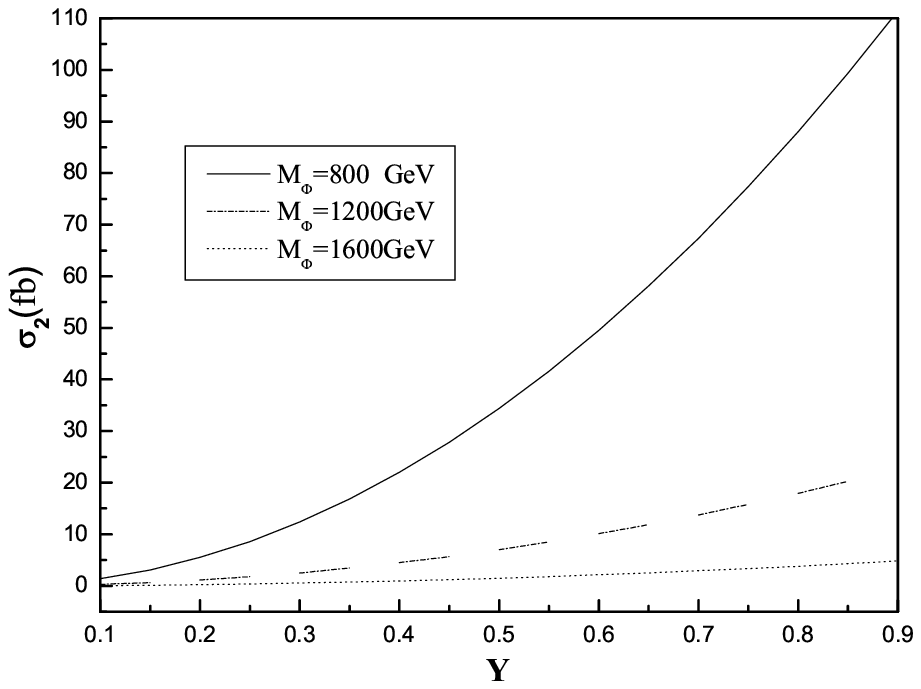}}
\end{picture}
\caption[]{The cross section $\sigma_{2}(s)$ as a function of the $FD$
coupling constant $Y$ for three values of the mass
$M_{\Phi}$.}
\label{fig4}
\end{center}
\end{figure}

The effective cross section $\sigma_{2}(s)$ of the process
$e^{-}p\rightarrow e^{+}X\Phi^{--}$  is plotted in Fig.4 as a
function of the $FD$ Yukawa coupling constant $Y$ for
$\sqrt{s}=3.7$TeV and three values of the mass $M_{\Phi}$. One can
see from Fig.4 that the cross section for single production of the
doubly charged scalar $\Phi^{--}$ at the $THERA$ is generally
smaller than that at the $ILC$ in most of the parameter space of the
$LH$ model. For $0.1\leq Y\leq0.9$ and $800GeV\leq M_{\Phi}\leq
1600GeV$, the value of the production cross section $\sigma_{2}(s)$
is in the range of $1.11\times10^{2}fb\sim5.9\times10^{-2}fb$. There
will be several tens of the doubly charged scalar $\Phi^{--}$ to be
generated per year at the $THERA$ with $\sqrt{s}=3.7$TeV and
$\pounds=100pb^{-1}$.

Similar to single production of $\Phi^{--}$ at the $ILC$, the
process $e^{-}p\rightarrow e^{+}X\Phi^{--}$ gives rise to number of
signal events with same-sign lepton pair and an isolated positive
electron in the $l^{-}l^{-}+e^{+}$ signature, which is almost free
of the $SM$ backgrounds. Thus, possible signatures of the doubly
charged scalar $\Phi^{--}$ might be detected in future $THERA$
experiments.

\section*{V. Single production of the doubly charged
scalar $\Phi^{--}$ at the $\emph{\textbf{LHC}}$}

\hspace{5mm}In this year, the $LHC$ with $\sqrt{s}=14$TeV and
$\pounds=100fb^{-1}$ will begin operation , which has an increase of
a factor of seven in energy and a factor of 100 in luminosity over
the $Fermilab$  $Tevatron$. The $LHC$ is expected to directly probe
possible $NP$ beyond the $SM$ up to a few TeV, which might provide
some striking evidence of $NP$.

\begin{figure}[t]
\setlength{\unitlength}{1mm}
\begin{center}
\begin{picture}(0,40)(0,0)
\put(-82,-150){\includegraphics{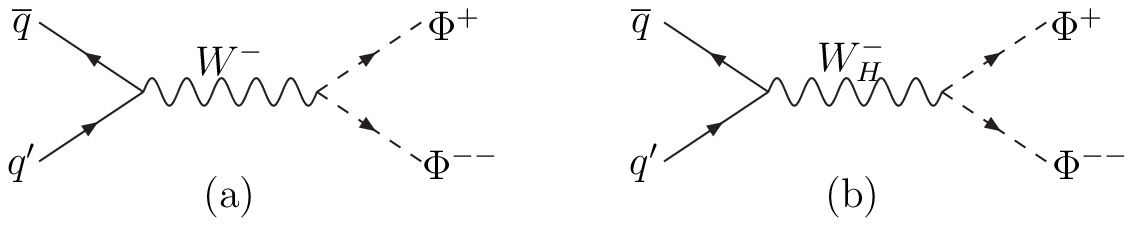}}
\end{picture}
\caption[]{Feynman diagrams for the partonic process $\overline{q}q'
 \rightarrow \Phi^{+}\Phi^{--}$.}
\label{fig5}
\end{center}
\end{figure}

From Eqs.(2) --- (8) given in section II, we can see that the doubly
charged scalar $\Phi^{--}$ can be singly produced at the $LHC$  via
the partonic processes:
\begin{eqnarray}
&&\overline{q}q'\rightarrow W^{-\ast},
\hspace*{0.15cm}W_{H}^{-\ast}\rightarrow W^{+}\Phi^{--};\\&&
qq\rightarrow
W^{-*}W^{-*}q'q'\rightarrow\Phi^{--}q'q';\\
&&\overline{q}q'\rightarrow W_{H}^{-\ast},
\hspace*{0.15cm}W^{-\ast}\rightarrow
W_{H}^{+}\Phi^{--};\\&&\overline{q}q'\rightarrow
W^{-\ast},\hspace*{0.15cm} W_{H}^{-\ast}\rightarrow\Phi^{+}\Phi^{--}
\hspace*{0.3cm}(q,q'=u, d, c, or \hspace*{0.15cm}s).
\end{eqnarray}
However, the production cross sections of the partonic processes
(15), (16), and (17) are strongly suppressed by the factor
$\nu'^{2}(\nu'\leq 1\times10^{-5}GeV)$. All of their values are
smaller than $1\times10^{-8}fb$. So, in this section, we only
consider the partonic processes $\overline{q}q'\rightarrow
W^{-\ast},W_{H}^{-\ast}\rightarrow \Phi^{+}\Phi^{--}$, as shown in
Fig.5.

Using Eq.(7), Eq.(8), and other relevant Feynman rules, we can write
the scattering amplitude for the partonic process
$\overline{q}(P_{1})+q'(P_{2})\rightarrow\Phi^{+}(P_{3})+\Phi^{--}(P_{4})$:
\begin{eqnarray}
\hspace*{-0.9cm}M\hspace*{-0.4cm}&=&\hspace*{-0.3cm}-\frac{e^{2}}{\sqrt{2}S_{W}^{2}}
|V^{SM}_{\overline{q}q'}|
\overline{v}(P_{1})P_{L}\gamma^{\mu}u(P_{2})\frac{g_{\mu\nu}}
{(P_{1}+P_{2})^{2}-M^{2}_{W}}(P_{3}-P_{4})^{\nu}
\nonumber\\&&\hspace*{-0.5cm}-\frac{ie^{2}(c^{2}-s^{2})}{2\sqrt{2}S_{W}^{2}s^{2}}|
V^{SM}_{\overline{q}q'}|
\overline{v}(P_{1})P_{L}\gamma^{\mu}u(P_{2})\frac{g_{\mu\nu}}
{(P_{1}+P_{2})^{2}-M^{2}_{W_{H}}}(P_{3}-P_{4})^{\nu}.
\end{eqnarray}
The cross section $\sigma_{3}(s)$ for single production of the
doubly charged scalar $\Phi^{--}$ associated with a singly charged
scalar $\Phi^{+}$ at the $LHC$ with $\sqrt{s}=14$TeV can be obtained
by convoluting the production cross section
$\widehat{\sigma}_{3}(\widehat{s})$ of the partonic process
$\overline{q}q'\rightarrow \Phi^{+}\Phi^{--}$ with the quark
distribution functions$(PDF's)$:
\begin{equation}
\sigma_{3}(s)=\sum_{ij}\int_{\tau}^{1}dx_{1}\int_{\tau/x_{1}}^{1}dx_{2}f_{i/p}(x_{1},
\mu)f_{j/p}(x_{2},\mu)\widehat{\sigma}(ij\rightarrow\Phi^{+}\Phi^{--}).
\end{equation}
Where $i$ and $j$ stand for the partons (light quarks u, d, c, or
s), $\tau=4M_{\Phi}^{2}/s$ and $\widehat{s}=x_{1}x_{2}s$ (We have
assumed $M_{\Phi}=M_{\Phi^{--}}= M_{\Phi^{+}}$). In our numerical
calculation, we will use $CTEQ6L$ $PDF's$ [28] for the quark
distribution functions and assume that the factorization scale is of
order $M_{\Phi}/2$.

From Eqs.(19) and (20), we can see that the production cross section
$\sigma_{3}(s)$ is dependent on the free parameters $M_{\Phi}$,
$M_{W_{H}}$, and $c$. It is well known that global fits to the
electroweak precision data impose rather severe constraints on the
free parameters of the $LH$ model. However, if the $SM$ fermions are
charged under $U(1)_{1}\times U(1)_{2}$, the constraints can become
relaxed [29]. As numerical estimation, we will assume that the free
parameters $M_{\Phi}$, $M_{W_{H}}$, and $c$ are in the ranges of
$800GeV\sim1600GeV$, $1TeV\sim2TeV$, and $0.1\sim0.5$, respectively.

Our numerical results are summarized in Fig.6, in which we plot the
production cross section $\sigma_{3}(s)$ as a function of the mass
$M_{\Phi}$ for different values of the free parameter $M_{W_{H}}$.
Since the contributions of the $LH$ model to the process
$pp\rightarrow \Phi^{+}\Phi^{--}X$ mainly come from $W$ exchange,
the production cross section $\sigma_{3}(s)$ is insensitive to the
free parameter $c$. So, we have taken $c=0.3$ in Fig.6. One can see
from Fig.6 that the cross section for single production of the
doubly charged scalar $\Phi^{--}$ at the $LHC$ is much smaller than
that at the $ILC$ or the $THERA$. This is because, at tree level,
the doubly charged scalar $\Phi^{--}$ is singly produced at the
$LHC$ via the s-channel processes with highly virtual gauge boson
propagators. For $800GeV\leq M_{\Phi}\leq1600GeV$, $1TeV\leq
M_{W_{H}}\leq2TeV$, and $c=0.3$, the value of $\sigma_{3}(s)$ is in
the range of $7.4\times10^{-1}fb\sim1.3\times10^{-4}fb$. Then, there
will be several tens $\Phi^{+}\Phi^{--}$ events to be generated one
year at the $LHC$ with $\sqrt{s}=14TeV$ and $£\pounds=100fb^{-1}$.

\begin{figure}[t]
\setlength{\unitlength}{1mm}
\begin{center}
\begin{picture}(0,60)(0,0)
\put(-42,-0){\includegraphics{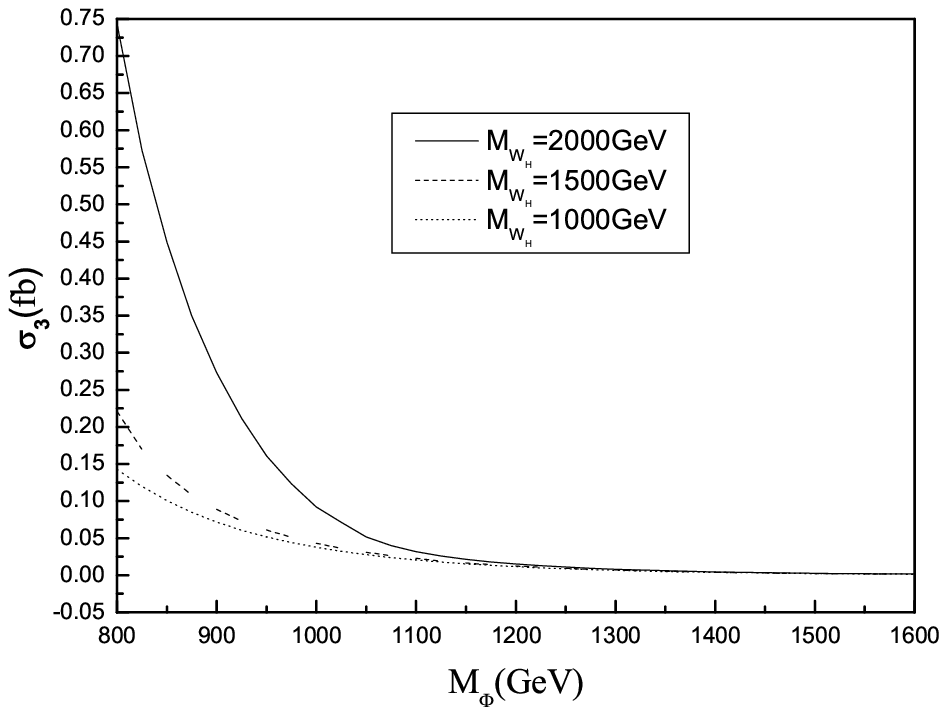}}
\end{picture}
\caption[]{The cross section $\sigma_{3}(s)$ as a function of the
scalar $\Phi$ mass $M_{\Phi}$ for three values of the
mass $M_{W_{H}}$.}
\label{fig6}
\end{center}
\end{figure}

The possible decay modes of the singly charged scalar $\Phi^{+}$ are
$l^{+}\overline{\nu}_{l}$, $t\overline{b}$, $T\overline{b}$,
$W^{+}Z$, and $W^{+}H$[15]. The decay widths
$\Gamma(\Phi^{+}\rightarrow W^{+}Z)$ and $\Gamma(\Phi^{+}\rightarrow
W^{+}H)$ are proportional to the factor $\nu'^{2}$, which can be
neglected. If we assume that the scalar $\Phi^{+}$ mass $M_{\Phi}$
is smaller than the vector-like top quark mass $M_{T}(M_{\Phi}\leq
M_{T})$, then the decay channel $\Phi^{+}\rightarrow T\overline{b}$
is kinematically forbidden. In this case, the singly charged scalar
$\Phi^{+}$ mainly decays into $l^{+}\overline{\nu}_{l}$ and
$t\overline{b}$. If the singly charged scalar $\Phi^{+}$ decays into
$l^{+}\overline{\nu}_{l}$, then the production of the doubly charged
scalar $\Phi^{--}$ associated with a singly charged scalar
$\Phi^{+}$ at the $LHC$ gives rise to a number of signal events with
like-sign di-leptons, with one jet and large missing energy,
$l^{-}l^{-}+\not{\hspace*{-0.2cm}E}+jet$, which has a relatively
high detection efficiency and enjoys essentially negligible
background from $SM$ processes. The production rate of the signal
event $l^{-}l^{-}+\not{\hspace*{-0.2cm}E}+jet$ is shown in Fig.7 as
a function of the $FD$ Yukawa coupling constant Y for $f=1$TeV and
$M_{\Phi}=1$TeV. One can see from Fig.7 that, for $0.1\leq Y\leq0.9$
and $1TeV\leq M_{W_{H}}\leq 2TeV$, the production rate of the signal
event is in the range of $6.7\times10^{-2}fb\sim1.4\times10^{-1}fb$.
With the high luminosity option of the $LHC$, around $300fb^{-1}$,
there will be several tens signal events to be generated one year.

\begin{figure}[t]
\setlength{\unitlength}{1mm}
\begin{center}
\begin{picture}(0,60)(0,0)
\put(-42,-0){\includegraphics{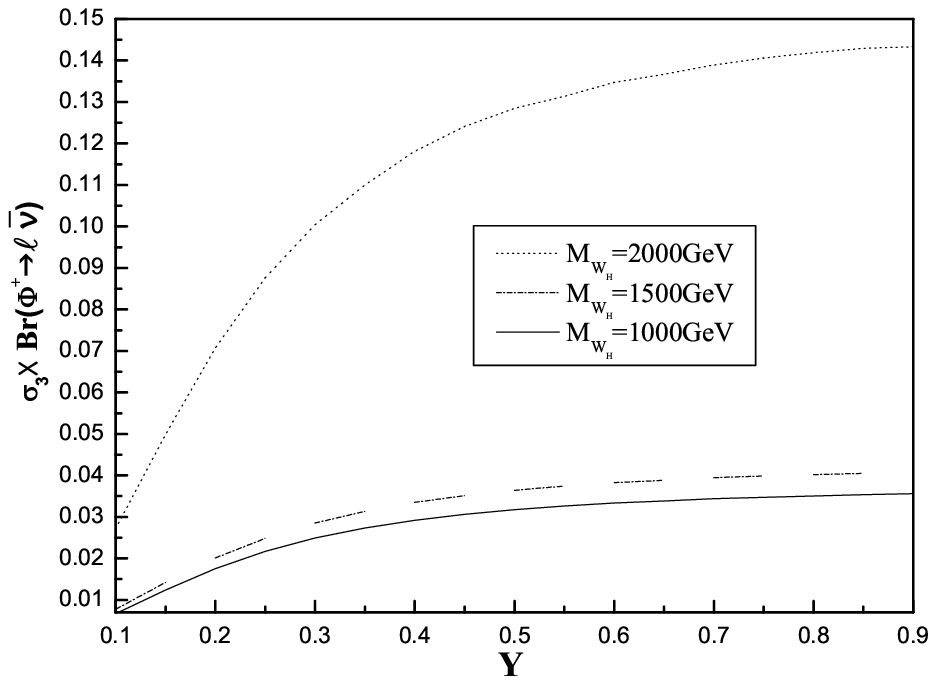}}
\end{picture}
\caption[]{Production rate of the signal event
$l^{-}l^{-}+\not{\hspace{-1.5mm}E}+jet$ as a function of the $FD$
coupling constant Y for three values of the $W_{H}$
mass $M_{W_{H}}$.}
\label{fig7}
\end{center}
\end{figure}

If we assume that the singly charged scalar $\Phi^{+}$ decays into
$t\overline{b}$, then the partonic process
$q\overline{q'}\rightarrow \Phi^{+}\Phi^{--}$ will generate the
$l^{-}l^{-}+\not{\hspace*{-0.2cm}E}+jets$ signature, which is
similar to that coming from production of the first or second
generation T-odd quark pair in the context of the $LH$ model with
T-parity[30]. However, the production rate of this kind of signal
events is too small to be detected in future $LHC$ experiments.
\section*{VI. Conclusions and discussions}
\hspace{5mm}Doubly charged scalars appear in some popular $NP$
models beyond the $SM$, motivated by their usefulness in generating
neutrino masses. This kind of new particles have distinct
experimental signals through their decay to same-sign lepton pairs.
Their observation in future high energy collider experiments would
be a clear evidence of $NP$ beyond the $SM$. Thus, searching for
doubly charged scalars is one of the main goals of future high
energy collider experiments.

Little Higgs models have generated much interest as possible
alternative to the $EWSB$ mechanism, which can be regarded as one of
the important candidates of $NP$ beyond the $SM$. As the most
economical little Higgs model, the $LH$ model can explain the
observed neutrino mass by introducing the lepton number violating
interaction of the triplet scalar to leptons. The neutrino mass is
proportional to the triplet scalar $VEV$ $\nu'$ multiplied by the
Yukawa coupling constant $Y_{ij}$ without invoking a right-handed
neutrino. This scenario predicts the existence of the doubly charged
scalars $\Phi^{\pm\pm}$. In this paper, we consider single
production of this kind of new particles in future three types of
high energy collider experiments( $ILC$, $THERA$, and $LHC$).

In future $ILC$ experiments, the doubly charged scalar $\Phi^{--}$
can be singly produced via $e^{-}\gamma$ collision. Our numerical
results show that it can be abundantly produced in future $ILC$
experiments. In most of the parameter space of the $LH$ model, there
will be a large number of the signal events with same-sign lepton
pair and an isolated positive electron to be generated in the $ILC$
experiment with $\sqrt{s}=2$TeV and $\pounds=100fb^{-1}$. Thus, the
possible signals of the doubly charged scalar $\Phi^{--}$ should be
detected in future $ILC$ experiments.

Using the $EPA$ method, we calculate the cross section for single
production of the doubly charged scalar $\Phi^{--}$ at the $THERA$
with $\sqrt{s}=3.7$TeV and $\pounds=100pb^{-1}$ via the subprocess
$e^{-}\gamma\rightarrow e^{+}\Phi^{--}$. In our numerical
estimation, we have included both contributions of  the elastic
photon and of the inelastic photon components from proton. We find
that, for $0.1\leq Y\leq0.9$ and $800GeV\leq M_{\Phi}\leq 1000GeV$,
the value of the production cross section $\sigma_{2}$ is in the
range of $1.11\times10^{2}fb\sim5.9\times 10^{-2}fb$. As long as the
doubly charged scalar $\Phi^{--}$ is not too heavy, its possible
signatures might be detected via the process $e^{-}p\rightarrow
e^{+}\Phi^{--}X$ in future $THERA$ experiments.

At the $LHC$, the doubly charged scalar $\Phi^{--}$ can be singly
produced via several patronic processes. However, considering that
the triplet scalar $VEV$ $\nu'$ is very small
$(\nu'\leq1\times10^{-5}GeV)$, the production channels involving the
gauge bosons $W$ or $W_{H}$ in the final states can be neglected.
Thus, in this paper, we only consider the production of the doubly
charged scalar $\Phi^{--}$ associated with a single charged scalar
$\Phi^{+}$ via the patronic processes $\overline{q}q'\rightarrow
W^{-*},W_{H}^{-*}\rightarrow\Phi^{+}\Phi^{--}$ at the $LHC$. Since
the partonic process $\overline{q}q'\rightarrow\Phi^{+}\Phi^{--}$
proceeds by the s-channel $W$ exchange and $W_{H}$ exchange, which
have highly virtual gauge boson propagators, its production cross
section is much smaller than that at the $ILC$ or the $THERA$.
However, the partonic process
$\overline{q}q'\rightarrow\Phi^{+}\Phi^{--}$ with
$\Phi^{+}\rightarrow l^{+}\nu_{l}$ and $\Phi^{--}\rightarrow
l^{-}l^{-}$ can produce distinct experimental signature, which is
almost free of the $SM$ background. If the mass of $\Phi^{--}$ is
smaller than 1TeV, its possible signals might be detected at the
high luminosity option of the $LHC$ experiments.

At the $ILC$, the doubly charged scalar $\Phi^{--}$ can also be
produced associated with a photon via $e^{-}e^{-}$ collision, i.e.
the subprocess $e^{-}e^{-}\rightarrow\Phi^{--}\gamma$. The
cleanliness of the final state photon detection in future
$e^{-}e^{-}$ collider can be very helpful in identifying the doubly
charged scalar. However, this production process suffers from a
large $SM$ background process
$e^{-}+e^{-}\rightarrow\gamma+e^{-}+e^{-}$, which has to be
carefully disposed by suitable technique [10].

Certainly, the doubly charged scalar $\Phi^{++}$ can also be singly
produced at the $ILC$, the $THERA$, and the $LHC$ via the
charge-conjugation processes of the corresponding processes for the
doubly charged scalar $\Phi^{--}$. Similar to above calculation, we
can give the values of the production cross sections for these
processes. Thus, the conclusions for the doubly charged scalar
$\Phi^{--}$ also apply to the doubly charged scalar $\Phi^{++}$.

The $LH$ model with T-parity also predicts the existence of the
doubly charged scalars $\Phi^{\pm\pm}$. While they have T-odd parity
and their $VEV$ is equal to zero, which can not generate the
neutrino masses via introducing the lepton number violating
interaction of the T-odd triplet scalar $\Phi$ to leptons. However,
they can also be produced associated a T-odd new particle in future
high energy collider experiments, which might be need to be further
studied.
 \vspace{0.5cm}

\noindent{\bf Acknowledgments}

This work was supported in part by Program for New Century Excellent
Talents in University(NCET-04-0290), the National Natural Science
Foundation of China under the Grants No.10475037 and 10675057.

\null
\end{document}